%% file: F-adjoints_a3.tex
\documentclass[12pt]{article}
\pdfoutput=1

\usepackage{amssymb,amsmath,bbm}
\usepackage{graphicx}
\usepackage{fullpage}
\usepackage{multirow}
\usepackage{color}
\usepackage{hyperref}
\usepackage{cite}

\input{macros.tex}

\numberwithin{equation}{section}

\begin{document}
\setcounter{page}{0}
\thispagestyle{empty}

\begin{center}
{\LARGE Dirac gauginos and unification in F-theory\\}
\vspace{.5in}
Rhys Davies\footnote{\it daviesr@maths.ox.ac.uk} \\
\vspace{.15in}
{\it
Mathematical Institute, \\
University of Oxford, \\
24-29 St Giles, Oxford \\
OX1 3LB, UK}
\end{center}

\begin{abstract}
Supersymmetric models in which the gauginos acquire Dirac masses,
rather than Majorana masses, offer an appealing alternative to the
minimal supersymmetric standard model, especially in the light of
the bounds set on superpartner masses by the 2011 LHC data.  Dirac
gauginos require the presence of chiral multiplets in the adjoint representation
of the gauge group, and the realisation of such scenarios in F-theory
is the subject of this paper.
The chiral adjoints drastically alter the usual picture of gauge coupling
unification, but this is disturbed anyway in F-theory models with
non-trivial hypercharge flux.  The interplay between these two factors
is explored, and it is found for example that viable F-theory unification
can be achieved at around the reduced Planck scale, if there is an extra
vector-like pair of singlet leptons with TeV-scale mass.
I then discuss the conditions which must be satisfied by the geometry
and hypercharge flux of an F-theory model with Dirac gauginos.  One
nice possibility is for the visible sector to be localised on a $K3$ surface,
and this is discussed in some detail.  Finally, I describe how to achieve
an unbroken discrete $R$-symmetry in such compactifications, which
is an important ingredient in many models with Dirac gauginos, and
write down a simple example which has adjoint chiral multiplets, an
appropriate $R$-symmetry, and allows for viable breaking of $SU(5)$
by hypercharge flux.
\end{abstract}

\newpage


\section{Introduction and Motivation}\label{sec:intro}

The last few years have seen a surge of interest in trying to construct realistic
models of particle physics within F-theory \cite{Vafa:1996xn}, following
foundational work in \cite{Donagi:2008ca,Beasley:2008dc,Hayashi:2008ba,
Beasley:2008kw,Donagi:2008kj,Donagi:2009ra}.  Much work has been done on
model building and phenomenology from both a local
(\!\!\!\cite{Marsano:2008jq,Heckman:2008qt,Font:2008id,Heckman:2008qa,
Bouchard:2009bu,Heckman:2009mn,Conlon:2009qa,Cecotti:2009zf,
Conlon:2009qq,Dudas:2009hu,King:2010mq,Heckman:2010fh,Dudas:2010zb,
Leontaris:2010zd,Choi:2010gx,Dolan:2011iu,Choi:2011ua,Leontaris:2011pu,
Heckman:2011hu,Aparicio:2011jx,Callaghan:2011jj,Dolan:2011aq,
Camara:2011nj,Kawano:2011aa,Palti:2012aa
})
and global (\!\!\!\cite{Marsano:2009ym,Marsano:2009gv,Blumenhagen:2009yv,
Marsano:2009wr,Grimm:2009yu,Cvetic:2010rq,Chen:2010tp,Chen:2010ts,
Grimm:2010ez,Knapp:2011wk,Marsano:2011nn,Braun:2011zm,Krause:2011xj,
Grimm:2011fx})
point of view.
One feature common to all studies which have appeared so far is that the
ultimate aim has been to reproduce, at low scales, the physics of the minimal
supersymmetric standard model (MSSM), in some appropriate corner of its
parameter space.  Here I wish to begin discussing the embedding in F-theory
of a different class of models --- those in which the gauginos acquire Dirac
masses, as opposed to Majorana masses, after supersymmetry breaking.
This necessarily involves extending the light spectrum, since, by definition,
a Dirac fermion is two \emph{different} left-handed spinors combined to give
a single massive particle, whereas a Majorana fermion consists of only one.
In order to have Dirac gauginos in a supersymmetric theory, we must
therefore add to the MSSM spectrum, chiral multiplets in the adjoint
representation of the gauge group.  After SUSY breaking, the fermions from
these multiplets can pair up with the gauginos to form Dirac gauginos.

The quantum behaviour of Dirac gauginos was studied in \cite{Fox:2002bu},
where it was shown that they give only finite, positive radiative corrections
to the squared sfermion soft masses (to be compared with the
logarithmically-divergent contributions from Majorana gauginos).  This means,
in particular, that the gauginos can be taken to be significantly heavier than
the electroweak scale, without paying a price in fine-tuning.  The assumption
of heavy Dirac gauginos considerably weakens the current LHC bounds on
squark masses \cite{Heikinheimo:2011fk,Kribs:2012gx}, making this a viable
alternative to so-called `natural SUSY' models, in which only the third
generation squarks are light enough to be produced at the LHC (see, e.g.,
\cite{Dine:1993np,Papucci:2011wy,Craig:2012hc,Craig:2012di}).  Extensive
work has been done on developing and studying field theory models with
Dirac gauginos
\cite{Benakli:2008pg,Benakli:2009mk,Benakli:2010gi,Carpenter:2010as,
Chun:2010hz,Abel:2011dc,Benakli:2011kz},
and a useful overview of their properties is given in \cite{Benakli:2011vb}.

There is another good reason to be interested in Dirac gauginos: they are
the only possibility in models with an unbroken (approximate)
$R$-symmetry\footnote{It is not necessary to consider $U(1)_R$ invariance;
the same conclusions follow from invariance under the $\IZ_p$ subgroup,
for $p>3$.} at the weak scale
\cite{Hall:1990hq,Nelson:2002ca,Kribs:2007ac,Davies:2011mp,
Frugiuele:2011mh,Unwin:2011ag,Rehermann:2011ax,Davies:2011js,Bertuzzo:2012su}.
This is because gauginos carry $R$-charge 1, and so cannot have Majorana masses
in the presence of unbroken $R$-symmetry.  $R$-symmetric models have the
appealing property that they have far fewer soft parameters than the MSSM,
and are safer from constraints on flavour and $CP$-violation
\cite{Kribs:2007ac}.  The extra symmetry also forbids dimension four and
five operators which can lead to proton decay.  

We will consider the standard F-theory GUT setup: F-theory compactified
on a Calabi--Yau fourfold $X$, elliptically-fibred over a K\"ahler threefold
$B$.  $X$ will be taken to have an $A_4$ singularity fibred over a complex
surface $S \subset B$; physically, this corresponds to having a stack of
branes wrapping $S$, which support an $SU(5)$ gauge theory.  I will
refer to this stack of branes as the `GUT brane'.  $SU(5)$ will be broken
to the standard model gauge group
$\GSM \equiv SU(3){\times}SU(2){\times}U(1)_Y$ by turning on a
non-trivial hypercharge flux on $S$, an approach pioneered in
\cite{Beasley:2008kw,Donagi:2008kj}.  The only major difference
between this work and all that which has preceded it is that the
geometry and flux will be chosen so that the theory contains light chiral
multiplets in the adjoint representation of $\GSM$.  Since techniques to
engineer a realistic matter sector have been developed at length in the
references, and should translate largely unchanged to this new context, I
will focus on this extended `adjoint sector'.

One dramatic consequence of the new chiral adjoint multiplets is that they
spoil the famous unification of the gauge coupling constants in the MSSM.
In four-dimensional theories, this makes it necessary to add quite a large
number of extra charged fields, in incomplete $SU(5)$ multiplets, if one
wishes to preserve unification \cite{Benakli:2010gi}.  In F-theory
with hypercharge flux breaking of $SU(5)$, equality of all three gauge
couplings at the unification scale is replaced by only a single linear
condition, first written down in \cite{Blumenhagen:2008aw}.  As this is
somewhat model independent, it is discussed first, in \sref{sec:unification},
with one notable conclusion being that in models with Dirac gauginos,
F-theory unification can be achieved at around the reduced Planck scale if we
also add just one vector-like pair of singlet leptons with TeV-scale mass.
Section \ref{sec:flux} then explains how to arrange for the presence of light
chiral adjoints in F-theory models.  After some generalities, I specialise to
the case where the GUT brane wraps a complex surface $S \cong K3$,
which has a number of nice features, and write down a necessary and
sufficient condition for the hypercharge flux to remove all massless fields
coming from the unwanted components of the adjoint of $SU(5)$.  In
\sref{sec:R_symmetry}, I consider engineering a discrete $R$-symmetry,
which must come from a geometric symmetry of the compactification.
$R$-symmetric Dirac masses require that the adjoint superfields have
$R$-charge zero, and the corresponding geometric condition is found
explicitly, again in the $K3$ case.  In \sref{sec:toy_model}, a simple example
is given of a compactification in which $SU(5)$ can be appropriately broken
by hypercharge flux, and there is an unbroken $\IZ_4$ $R$-symmetry under
which the adjoint fields have the correct charge.  Section
\ref{sec:conclusion} briefly concludes.  There is also a short appendix
reviewing some of the field-theoretic issues associated with generating
Dirac gaugino masses.

\section{Unification}\label{sec:unification}

One attractive feature of the MSSM is that the three standard model gauge
couplings unify to good precision at
$M_\GUT \simeq 2.1\times 10^{16} \GeV$.  Since the new adjoint fields we
wish to introduce do not fill out complete multiplets of $SU(5)$, their presence
spoils this unification, forcing us to add further new charged states, although
the situation in F-theory turns out to be somewhat less restrictive than in
four-dimensional GUTs.  Throughout this section, I will consider only
one-loop running, and neglect threshold corrections, so all conclusions are
somewhat qualitative.

For a gauge theory with coupling constant $g$, we define the fine structure
constant $\a = \frac{g^2}{4\p}$; the one-loop relation between the values of
this quantity measured at two energy scales $\mu$ and $\L$ is
\begin{equation}\label{eq:running_coupling}
    \a^{-1}(\mu) = \a^{-1}(\L) + \frac{b}{2\p} \log\left(\frac{\m}{\L}\right) ~,
\end{equation}
where $b$ is a constant\footnote{Some authors use a convention in which
$b$ has the opposite sign.} to which each charged field in the theory
contributes additively.

The MSSM has three independent couplings for $U(1)_Y$, $SU(2)$, and
$SU(3)$, which we can denote as $g_Y, g_2, g_3$ respectively.  We will
take the generator of $U(1)_Y$ to be embedded in $\mathfrak{su}(5)$ as
$T_Y = \diag(-2,-2,-2,3,3)$ in the fundamental
representation,\footnote{Traditionally, the hypercharge generator is taken
to be $\frac13 T_Y$ or $\frac16 T_Y$; our normalisation is chosen so that
$e^{2\pi\ii T_Y} = \id$.} but to study unification, we must use an appropriately
rescaled version. The generators of $SU(n)$ are conventionally taken to
satisfy $\Tr\,T^2 = \frac12$ in the fundamental representation, so we define
$T_1 = \frac{1}{\sqrt{60}}T_Y$; it is the corresponding coupling constant
$g_1$ which we should compare to $g_2$ and $g_3$.   In the MSSM, the
one-loop beta function coefficients (above all mass scales in the theory)
are then
\begin{equation*}
    b_1 = -\frac{33}{5} ~,~~ b_2 = -1 ~,~~ b_3 = 3 ~.
\end{equation*}

\begin{figure}
\begin{center}
    \includegraphics[width=.75\textwidth]{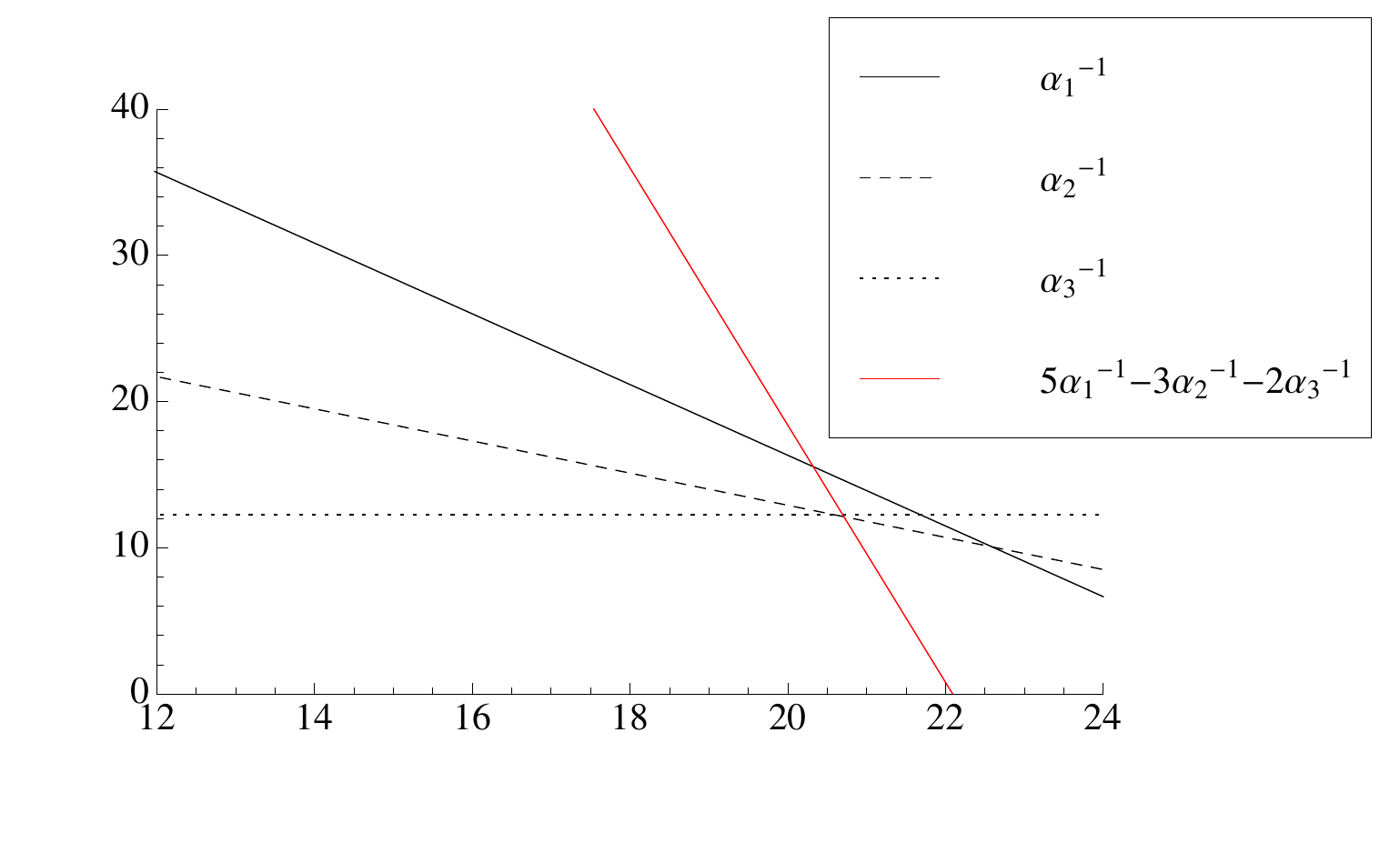}\\
    \place{4.8}{0.6}{$\log_{10}(\frac{E}{\mathrm{GeV}})$}
    \vspace{-6ex}
    \parbox{\textwidth}{
    \caption{\label{fig:adjoint_running}
        \small
        One-loop running of the supersymmetric standard model gauge coupling constants
        at high energies, in the presence of adjoint chiral multiplets.  In this plot, the (Dirac)
        gauginos and adjoint scalars have masses of $5\TeV$, while all other non-standard
        model states have masses of $1\TeV$.  We see that the relation
        \eqref{eq:F_unification}, which defines the GUT scale, only holds well above the
        Planck scale.}}
\end{center}
\end{figure}

Now consider the situation with light chiral adjoints.  A chiral multiplet in
the adjoint of $SU(n)$ makes a contribution $\d b = -n$, so we find
\begin{equation*}
    \d b_1 = 0 ~~,~ \d b_2 = -2 ~~,~ \d b_3 = -3 ~.
\end{equation*}
The different contributions to the three couplings mean that, if we assert
their measured values at the $Z$ pole, then they no longer unify at any scale,
as can be seen in \fref{fig:adjoint_running}.  Note also that now
$b_3 = 0$, i.e., the $SU(3)$ coupling constant no longer runs at high scales.

In F-theory, however, the presence of the chiral adjoint fields is not the
only thing which interferes with standard gauge unification.  Although
the standard model gauge fields all arise from an underlying $SU(5)$,
turning on hypercharge flux to break this to $\GSM$ can already cause
a discrepancy between their coupling constants at tree level
\cite{Donagi:2008kj,Blumenhagen:2008aw}.  The details of flux breaking
will be discussed in \sref{sec:flux}, but for now it suffices to say that it
involves a choice of two line bundles $\cL_a$ and $\cL_Y$ on the complex
surface $S$ on which the $SU(5)$ theory lives, and leads to the following
expressions \cite{Blumenhagen:2008aw}:\footnote{There are also extra
terms which are sub-dominant at weak-coupling \cite{Grimm:2012rg}.}
\begin{equation} \label{eq:tree_level_couplings}
    \begin{array}{r c l}
        \a_1^{-1} & = & \a_{\mathrm{YM}}^{-1}\Vol(S) - \frac{1}{2g_s} \int_S \left(c_1(\cL_a)^2 + \frac 65 c_1(\cL_a)c_1(\cL_Y)
            + \frac{3}{5} c_1(\cL_Y)^2\right) \\[2ex]
        \a_2^{-1} & = & \a_{\mathrm{YM}}^{-1}\Vol(S) - \frac{1}{2g_s} \int_S \big(c_1(\cL_a)^2 +
            2c_1(\cL_a)c_1(\cL_Y) + c_1(\cL_Y)^2\big) \\[2ex]
        \a_3^{-1} & = & \a_{\mathrm{YM}}^{-1}\Vol(S) - \frac{1}{2g_s} \int_S c_1(\cL_a)^2 ~,
    \end{array}
\end{equation}
where $\a_{\mathrm{YM}}$ is the 8D Yang-Mills coupling, and $g_s$ is the
string coupling.  The three coupling constants therefore depend (differently)
on the choice of the line bundles $\cL_Y$ and $\cL_a$, and the intersection
form on $S$, but there is one invariant relation:
\begin{equation}\label{eq:F_unification}
    5\a_1^{-1} - 3 \a_2^{-1} - 2 \a_3^{-1} = 0 ~.
\end{equation}
This is the relationship that holds between the couplings at the compactification
scale.\footnote{The couplings will all be equal, as in traditional GUTs, if
$\int_S \big(2c_1(\cL_a)c_1(\cL_Y) + c_1(\cL_Y)^2\big) = 0$.}  Given the low
energy spectrum, we find the compactification/GUT scale by running the couplings
up until \eqref{eq:F_unification} is obeyed.

If we consider just the MSSM spectrum, augmented by light chiral adjoint fields,
then \eqref{eq:F_unification} is only obeyed well above the Planck scale (see
\fref{fig:adjoint_running}), where the calculation no longer makes sense.  As such,
this scenario is ruled out, and we are forced to consider the addition of further light
multiplets.

\subsection{Extra vector-like matter}

Which extra vector-like states can we add to bring the GUT scale down to
something realistic?  First note that at one loop, particles contribute additively
to $\a^{-1}$; a state of mass $M$ changes the value of $\a^{-1}$ at scales
$\mu > M$ by an amount
\begin{equation*}
    \d\a^{-1} = \frac{\d b}{2\pi}\log\left(\frac{\mu}{M}\right) ~. 
\end{equation*}
To study the effect of the spectrum on the GUT scale, defined by
\eqref{eq:F_unification}, we must therefore consider the combination
$\d b_F := 5\,\d b_1 - 3\,\d b_2 - 2\,\d b_3$ for each possible multiplet: the values
are given in \tref{tab:beta_coefficients} (the dependence of the GUT scale
on extra vector-like matter, as well as threshold corrections, has also been
discussed in \cite{Leontaris:2009wi,Leontaris:2011tw}).

We can see that
the $\GSM$ adjoints make a total contribution of $\d b_F = 12$, which is what
raises $M_\GUT$ from its usual value to well above the Planck scale.  To
bring it back down, we must introduce light states with $\d b_F < 0$.
The minimal possibility is to add one light ($\sim 1\TeV$) vector-like pair of
singlet leptons,\footnote{We will discuss in \sref{sec:flux} how this might be
arranged in F-theory.} i.e., chiral multiplets in the representation
$(\rep{1},\rep{1},6)\oplus(\rep{1},\rep{1},-6)$.  The effect of this addition is to
bring $M_\GUT$ back down to $M_\GUT \simeq 1.7 \times 10^{18}\GeV$,
which is approximately the reduced Planck scale
$M_P \simeq 2.4 \times 10^{18}\GeV$.  The usual small hierarchy between
the Planck scale and the GUT scale is therefore removed in this scenario,
resulting in a complete unification.  Of course, this is not a firm prediction: if the
mass and charges of the extra vector-like states are varied, then there are
a number of ways to bring $M_\GUT$ to a reasonable value, with the only
point of general concern being that all couplings remain perturbative, so that
the calculations can be trusted.
\begin{table}[ht]
\def\str{\vrule height3ex width0pt depth2ex}
\begin{center}
    \begin{tabular}{| c | c | c | c | c | c |}
        \hline
        $SU(5)$ irrep. & $\GSM$ irrep. & ~$\d b_1$~ & ~$\d b_2$~ & ~$\d b_3$~ & ~$\d b_F := 5\,\d b_1 - 3\,\d b_2 - 2\,\d b_3$~ \str \\
        \hline
        \multirow{3}{*}{\raisebox{-3ex}{$\rep{10}$}}
          & $(\conjrep{3},\rep{1},-4)$ & $-\frac45$ & $0$ & $-\frac12$ & $-3$
        \str\\
          & $(\rep{3},\rep{2},1)$ & $-\frac{1}{10}$ & $-\frac32$ & $-1$ & $\+6$
        \str\\
          & $(\rep{1},\rep{1},6)$ & $-\frac35$ & $0$ & $0$ & $-3$ \str\\
        \hline 
        \multirow{3}{*}{\raisebox{1ex}{$\conjrep{5}$}}
          & $(\conjrep{3},\rep{1},-2)$ & $-\frac15$ & $0$ & $-\frac12$ & $\+0$ \str\\
          & $(\rep{1},\rep{2},3)$ & $-\frac{3}{10}$ & $-\frac12$ & $0$ & $\+0$ \str\\
        \hline
        \multirow{5}{*}{\raisebox{-6ex}{$\rep{24}$}} & $(\rep{8},\rep{1},0)$ & $-3$ & $0$ & $0$ & $\+6$ \str \\
            & $(\rep{1},\rep{3},0)$ & $0$ & $-2$ & $0$ & $\+6$ \str\\
            & $(\rep{1},\rep{1},0)$ & $0$ & $0$ & $0$ & $\+0$ \str\\
            & $(\rep{3},\rep{2},-5)$ & $-\frac52$ & $-\frac32$ & $-1$ & $-6$ \str \\
            & $(\conjrep{3},\rep{2},5)$ & $-\frac52$ & $-\frac32$ & $-1$ & $-6$ \str \\
        \hline
    \end{tabular}
    \parbox{\textwidth}{
        \caption{\label{tab:beta_coefficients}
        \small
        Contributions of chiral multiplets in the relevant standard model representations
        to the one-loop beta function coefficients.  Note that a vector-like pair of any of
        these will make twice the given contribution.
        }
    }
\end{center}
\end{table}

\vspace{-3ex}
The other interesting feature to notice in \tref{tab:beta_coefficients} is that
the $\GSM$ representations originating in the $\rep{5}{\oplus}\conjrep{5}$
of $SU(5)$ each have $\d b_F = 0$, so the presence of such fields does not
change $M_\GUT$.  We may therefore introduce any such states as
messengers of SUSY breaking, or as extra Higgs doublets (as required in
the `Minimal $R$-Symmetric Supersymmetric Model (MRSSM) \cite{Kribs:2007ac}),
without changing $M_\GUT$.  Such states will have the effect of making
the theory more strongly coupled at $M_\GUT$, but this is not necessarily
a problem.  For example, even in the MRSSM with an extra vector-like pair
of singlet leptons, we can add messengers filling out two full copies of
$\rep{5}\oplus\conjrep{5}$, with masses as low as $\sim 10^{12}\GeV$, and
the largest coupling constant at $M_\GUT$ is $\alpha_2 \simeq \frac 15$, so
the theory (just) remains perturbative.

\begin{figure}
\begin{center}
    \includegraphics[width=.45\textwidth]{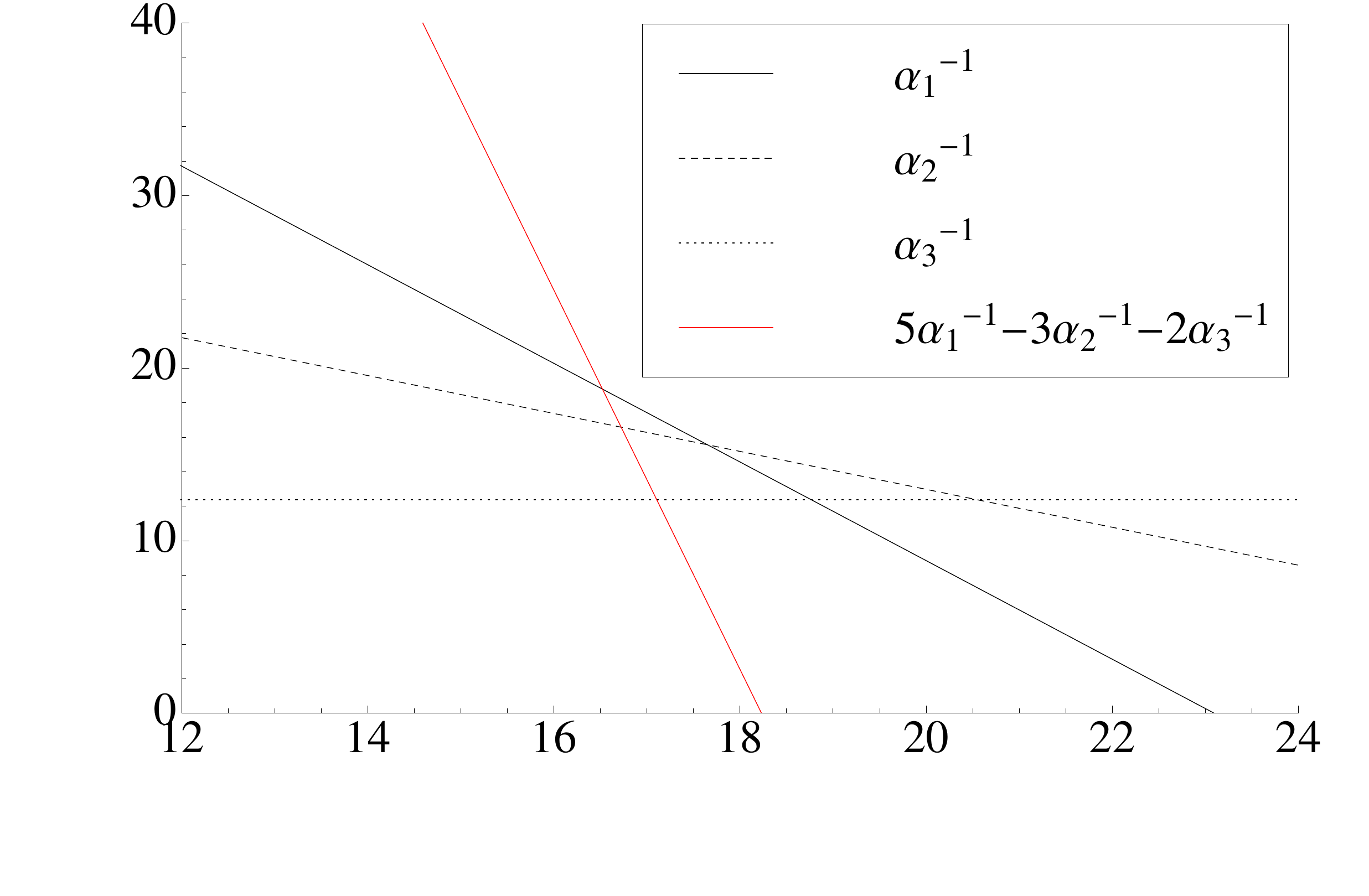}
    \includegraphics[width=.45\textwidth]{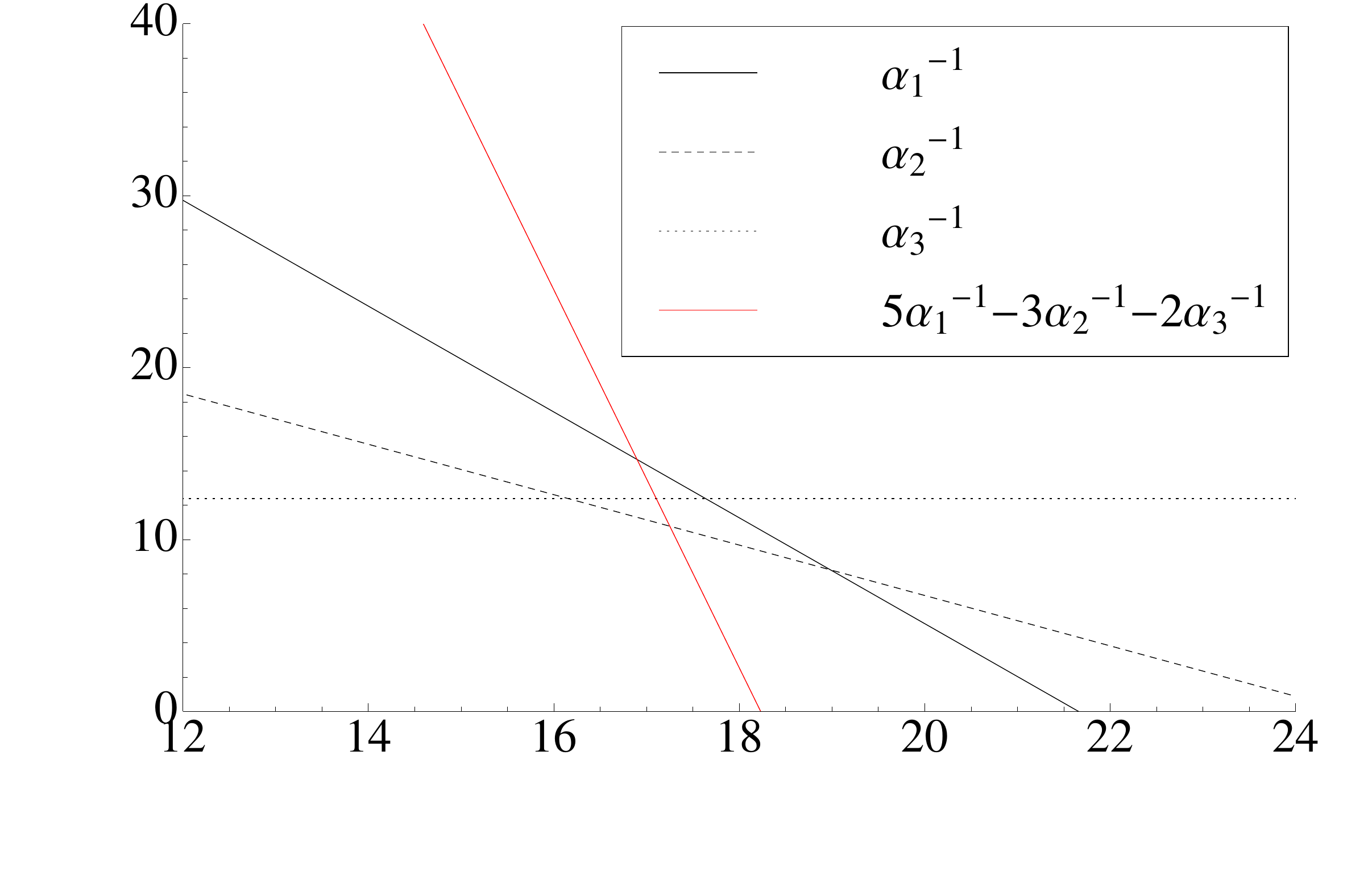}\\
    \place{2.85}{0.4}{$\log_{10}(\frac{E}{\mathrm{GeV}})$}
    \place{5.8}{0.4}{$\log_{10}(\frac{E}{\mathrm{GeV}})$}
    \vspace{-4ex}
    \parbox{\textwidth}{
    \caption{\label{fig:positron_running}
        \small
        These plots show the running coupling constants in two different
        theories, each with a single light ($1\TeV$) vector-like pair of
        particles in the $(\rep{1},\rep{1},\pm 6)$ representation of $\GSM$.  In
        the first, the rest of the spectrum is the same as in
        \fref{fig:adjoint_running}, while in the second there is an extra
        pair of Higgs doublets, as required in the MRSSM.  For the sake of
        the plot, these extra doublets are also given masses of $1\TeV$.  In each
        case, the couplings `unify', in the F-theory sense, at $\sim 1.7 \times 10^{18}\GeV$,
        which is just below the reduced Planck scale.
        }}
\end{center}
\end{figure}

One final point to note is the order of the couplings at $M_\GUT$.  The
relation in \eqref{eq:F_unification} can be re-written as
\begin{equation*}
   5(\a_1^{-1} - \a_3^{-1}) = 3(\a_2^{-1} - \a_3^{-1}) ~,
\end{equation*}
but does not say anything about the sign of these quantities; we
see from \eqref{eq:tree_level_couplings} that the sign is opposite to that of
$\int_S \big(2c_1(\cL_a)c_1(\cL_Y) + c_1(\cL_Y)^2\big)$.  As we will see in
\sref{sec:flux}, the simplest choice of flux, $c_1(\cL_a) = 0$,
makes this quantity negative, leading to the GUT-scale relation
$\a_3^{-1} < \a_1^{-1} < \a_2^{-1}$.  As shown in \fref{fig:positron_running},
this is obeyed in the case where the adjoint chiral fields and vector-like
leptons are the only addition to the MSSM spectrum, whereas the other
possibility ($\a_2^{-1} < \a_1^{-1} < \a_3^{-1}$) occurs if we also add an
extra light pair of Higgs doublets (as in the MRSSM).  Note that complete
$SU(5)$ multiplets do not affect such relations.

%

\section{Hypercharge flux and massless chiral adjoints}\label{sec:flux}

Our starting assumption is that the complex surface $S$ is wrapped by a stack
of branes which give rise to an $SU(5)$ gauge theory in eight dimensions.  As
has been common in the literature since the pioneering work in
\cite{Beasley:2008kw,Donagi:2008kj}, we will then break $SU(5)$ to $\GSM$
by turning on a non-trivial hypercharge flux\footnote{It is also possible to consider
discrete Wilson line breaking, as in heterotic models \cite{Braun:2010hr}.} along
$S$.  In order to preserve supersymmetry, the field strength $F_Y$ representing
this flux must be of Hodge type $(1,1)$, thus corresponding to some holomorphic
line bundle $\cL_Y$, and satisfy
\begin{equation} \label{eq:primitive_flux}
    F_Y\wedge \o_S = 0 ~,
\end{equation}
where $\o_S$ is the K\"ahler form on $S$.  We must also ensure that the
hypercharge gauge boson remains massless; this will be the case if
$c_1(\cL_Y) = \frac{1}{2\p}[F_Y]$ pushes forward to zero in the cohomology of
$B$.  The dual picture is often more convenient: $\cL_Y$ corresponds to
some divisor $D_Y$, which is a linear combination of algebraic
curves on $S$, and the hypercharge gauge boson remains massless if this is
homologous to zero in $B$.

We wish to engineer models which contain massless chiral multiplets in the
adjoint representation of the gauge group, and there are two possible
sources of the scalars in these multiplets: internally-polarised zero modes
of the eight-dimensional gauge fields, and zero modes of the
eight-dimensional scalar $\varphi$.  Since the flux breaks $SU(5)$ to $\GSM$,
we must decompose the adjoint representation into irreducible
representations of $\GSM$, and consider each separately:
\begin{equation}\label{eq:SU(5)_adjoint}
    \begin{array}{r c l}
        SU(5) & \supset & SU(3){\times}SU(2){\times}U(1)_Y \\[1ex]
        \rep{24} & = & (\rep{8}, \rep{1}, 0)\oplus(\rep{1}, \rep{3}, 0)\oplus(\rep{1}, \rep{1}, 0)\oplus
            (\rep{3}, \rep{2}, -5)\oplus(\conjrep{3}, \rep{2}, 5) ~,
    \end{array}
\end{equation}
The gauge bundle correspondingly splits into a
sum of line bundles of the form $\cL_Y^k$, where $k$ is the hypercharge
of the corresponding field components.  As discussed in
\cite{Donagi:2008ca,Beasley:2008dc}, the massless chiral multiplets
descending from $\vph$ correspond to the cohomology
groups\footnote{$K_S$, $K_B$ etc. will be used to denote both the
canonical divisor (class), and the canonical line bundle.  It should be clear
from context which is meant.}
$H^0(S, K_S\otimes\cL_Y^k)$, while those descending from the gauge
fields correspond to $H^1(S, \cL_Y^k)$; the scalars in these multiplets
are respectively the 7-brane position moduli, and continuous Wilson line
moduli.

As we can see from \eqref{eq:SU(5)_adjoint}, the fields in the adjoint of
$\GSM$ all have hypercharge zero, so we get massless chiral adjoints
from the cohomology groups $H^0(S, K_S) \cong H^{2,0}(S)$ and
$H^1(S, \cO_S) \cong H^{0,1}(S)$.  The surface $S$ is K\"ahler, so its
Hodge numbers satisfy $h^{p,q} = h^{q,p}$; we therefore seek surfaces
with either $h^{1,0}$ or $h^{2,0}$ equal to one.

It is not clear at this stage whether we should prefer surfaces with
$h^{1,0} = 1$, or $h^{2,0} = 1$.  However, a particularly nice possibility
is to take $S \cong K3$, the Hodge diamond for which is well known:
\begin{equation*}
    \begin{matrix}
       ~~~ h^{0,0} ~~~ \\
       ~~ h^{1,0} ~~ h^{0,1} ~~ \\
       ~ h^{2,0} ~ ~ h^{1,1} ~ ~ h^{0,2} ~ \\
       ~ ~ h^{2,1} ~ ~ h^{1,2} ~ ~ \\
       ~ ~ ~ h^{2,2} ~ ~ ~ \\
       \end{matrix}\hspace{10pt}
    =
    \hspace{10pt}
    \begin{matrix}
       ~~~ 1 ~~~ \\
       ~~ 0 ~~~ 0 ~~ \\
       ~ 1 ~ ~ ~20~ ~ ~ 1 ~ \\
       ~ ~ 0 ~~~ 0 ~ ~ \\
       ~ ~ ~ 1 ~ ~ ~ \\
    \end{matrix}
\end{equation*}
\newpage
Although other surfaces are probably just as suitable, there are several reasons
to consider $S \cong K3$:
\begin{itemize}
    \item
    When $h^{2,0} > 0$, the Picard number of $S$ is generally smaller than
    $h^{1,1}(S)$, since the integral cohomology lattice $H^2(S,\IZ)$ need not align
    with the subspace of $(1,1)$-forms in $H^2(S, \IC)$.  However, as the
    complex structure of $S$ is deformed by moving it around in $B$, new divisor
    classes, not inherited from $B$, can appear.\footnote{This phenomenon has been
    used in \cite{Braun:2011zm} to construct $G$-flux in global F-theory models.}
    Turning on hypercharge flux corresponding to such a divisor leaves the
    hypercharge gauge boson massless, and also fixes some of the moduli.  This
    mechanism is not available on a surface with $h^{2,0} = 0$.
    \item
    The trivial canonical bundle of $K3$ simplifies many calculations
    involving Serre duality or adjunction.
    \item
    It is very easy to find $K3$ surfaces embedded in appropriate threefolds $B$:
    any smooth anti-canonical hypersurface will be a $K3$.
    \item
    The special case $B = K3{\times}\IP^1$ is dual to the heterotic string compactified on
    $K3{\times}T^2$.  This theory has $\cN=2$ supersymmetry, with the adjoint chiral
    multiplets combining with the gauge fields to give $\cN=2$ vector multiplets.
    The actual case of interest is that in which the global geometry of $B$ (and the
    flux) instead breaks this to $\cN = 1$, but we see that $S \cong K3$ nicely realises
    the idea from \cite{Fox:2002bu} of having $\cN=2$ SUSY in the gauge sector only.
\end{itemize}
For these reasons we will frequently return to the case $S \cong K3$.

\subsection{Absence of unwanted states}

Demanding that $h^{1,0}(S) + h^{2,0}(S) = 1$ guarantees that the theory will
have massless chiral multiplets filling out exactly one copy of the adjoint of
$\GSM$, regardless of the hypercharge flux.  We now demand that there are
no light multiplets in the `off-diagonal' representations appearing in the
decomposition of the $SU(5)$ adjoint, \eqref{eq:SU(5)_adjoint}.  These carry
five units of hypercharge, so according to the discussion in the last section,
the necessary conditions are
\begin{equation*}
    H^0(S, K_S{\otimes}\cL_Y^{\pm 5}) = H^1(S, \cL_Y^{\pm 5}) = 0 ~.
\end{equation*}
As has been pointed out many times in the literature, and as we will see
explicitly below, this proves impossible to satisfy in cases of interest.  The
solution to this problem is to consider a slightly different type of flux.  We suppose
that our $SU(5)$ gauge group is in fact embedded in $U(5)$ (in what sense this
might be true depends on global features of the compactification).  The global
structure of $U(5)$ is in fact
\begin{equation}\label{eq:U(5)}
    U(5) = \frac{SU(5){\times}U(1)_a}{\IZ_5} ~,
\end{equation}
where $U(1)_a$ is the central `diagonal' subgroup.  Let $T_Y$ be the generator
of hypercharge $U(1)_Y$, and $T_a$ be the generator of $U(1)_a$, and take the
field strength of the line bundle $\cL_Y$ to correspond to $\frac{1}{5}(T_Y + 2T_a)
= \diag(0,0,0,1,1)$.  Despite being a fractional linear combination, this is an
appropriately normalised $U(1)$ generator, thanks to the global identification in
\eqref{eq:U(5)}.  It is easy to see that the charges of the off-diagonal components
in \eqref{eq:SU(5)_adjoint} with respect to this new $U(1)$ are simply $\pm 1$,
so the conditions for the absence of exotics become
$H^0(S, K_S{\otimes}\cL_Y^{\pm 1}) = H^1(S, \cL_Y^{\pm 1}) = 0$, and we will
see that these are easy to satisfy.

The first thing to note is that Serre duality gives us an isomorphism
\begin{equation*}
    H^0(S, K_S{\otimes}\cL_Y^{\pm 1}) = H^2(S, \cL_Y^{\mp 1})^*,
\end{equation*}
so our conditions can be recast as $H^i(S, \cL_Y^{\pm 1}) = 0$
for $i=1,2$.  At this point it is useful to introduce the holomorphic Euler
characteristic, given by
$\chi(S, \cL_Y^{\pm 1}) = \sum_{i=0}^2 (-1)^i h^i(S, \cL_Y^{\pm 1})$.  This can
be calculated easily, but will only be useful if we know the value of
$h^0(S, \cL_Y^{\pm 1})$.  Assuming that \eqref{eq:primitive_flux} holds, it is
in fact easy to show that $H^0(S,\cL_Y^{\pm 1}) = 0$.  To see this, note that if
$\cL_Y^k$, for any $k\neq 0$, were to admit a global section, it would mean
that $k\,c_1(\cL_Y) = \frac{k}{2\p}[F_Y]$ was dual to an algebraic curve
$C \subset S$, and therefore
\begin{equation*}
    \int_S F_Y\wedge \o_S = \frac{2\p}{k} \int_C \o_S = \frac{2\p}{k} \Vol(C) \neq 0 ~.
\end{equation*}
We conclude that $F_Y\wedge \o_S = 0$ is sufficient to ensure that
$H^0(S, \cL_Y^k) = 0$ for all $k \neq 0$, and therefore
$\chi(S, \cL_Y^{\pm 1}) = -h^1(S, \cL_Y^{\pm 1}) + h^2(S, \cL_Y^{\pm 1})$.

On a complex surface $S$, the Hirzebruch-Riemann-Roch theorem gives
the following formula for the holomorphic Euler characteristic:
\begin{align*}
    \chi(\cL_Y^{\pm 1}) ~&= \chi(\cO_S) + \frac{1}{2}\big(D_Y^2 \pm D_Y\cdot K_S\big) \\[1ex]
        &=  \chi(\cO_S) + \frac{1}{2}D_Y^2 ~.
\end{align*}
The second equality here follows from our assumption that $D_Y$ is
homologically trivial in $B$, whereas by the adjunction formula, $K_S$ is
the restriction to $S$ of a divisor on $B$, namely $S + K_B$.  The first term
is given in terms of the Hodge numbers of $S$:
\begin{equation*}
    \chi(\cO_S) = h^{0,0}(S) - h^{1,0}(S) + h^{2,0}(S) = 
        \left\{ \begin{array}{l}
            0 ~~\mathrm{if}~~ h^{1,0}(S) = 1, h^{2,0}(S) = 0 \\
            2 ~~\mathrm{if}~~ h^{1,0}(S) = 0, h^{2,0}(S) = 1~. \end{array}\right.
\end{equation*}
We conclude that a necessary condition to project out the unwanted states
is that $D_Y^2 = 0$ in the first case, or $D_Y^2 = -4$ in the second.  We see
now why pure hypercharge flux cannot work in the case $h^{2,0} = 1$:
since the off-diagonal states have five units of hypercharge, the condition
would be $(5D_Y)^2 = -4$, which is impossible, since $D_Y$ is
an integral class.  This is what necessitates the discussion in terms of $U(5)$
which I gave above.

Interestingly, there is no such problem in the first case, as the
condition remains simply $D_Y^2 = 0$.  Although I will not explore this
further here, it shows that pure hypercharge flux might work on surfaces with
$h^{1,0} = 1$, potentially leading to some simplifications in building global
models.

There does not seem to be anything more we can say in complete generality,
but in the case $S \cong K3$, we can easily go further.  Since
$K_S \cong \cO_S$,
$H^0(S, K_S{\otimes}\cL_Y^{\pm 1}) = H^0(S, \cL_Y^{\pm 1})$, and we have
already seen that the latter vanishes for a supersymmetric compactification.
But now it follows from Serre duality that $H^2(S, \cL_Y^{\pm 1}) = 0$, so in
this case we have simply $\chi(\cL_Y^{\pm 1}) = -h^1(S, \cL_Y^{\pm 1})$, and
the condition $D_Y^2 = -4$ becomes both necessary and sufficient for the
vanishing of all cohomology groups.

So what are the possible choices for $D_Y$?  As explained above, neither
$D_Y$ nor $-D_Y$ can be effective, or \eqref{eq:primitive_flux} would be
violated.  Therefore we must write $D_Y \sim C_1 - C_2$, where each $C_j$ is
a curve on $S$ (we assume them to be irreducible for simplicity).  Then
$D_Y^2 = C_1^2 + C_2^2 - 2 C_1\cdot C_2$.  On a $K3$ surface, the
adjunction formula gives, for any curve $C$,
\begin{align} \label{eq:genus_formula}
    K_C = \big(K_S + C\big)\rest{C} ~&= C\rest{C} ~,\notag\\[1ex]
    \Rightarrow~ \deg K_C ~&= \deg C\rest{C} ~,\notag\\[1ex]
    \Rightarrow~ 2g_C - 2 ~&= C^2 ~,
\end{align}
where $g_C$ is the genus.  Given this identity, $D_Y^2 = -4$ becomes
\begin{equation*}
    g_{C_1} + g_{C_2} - C_1\cdot C_2 = 0 ~.
\end{equation*}
It is possible to arrange for cancellation to take place here, but the simplest
solution is clearly $g_{C_1} = g_{C_2} = C_1 \cdot C_2 = 0$, i.e., $C_1$ and
$C_2$ are disjoint rational curves on $S$.\footnote{Taking $C_1$ and $C_2$
to be disjoint $(-2)$-curves will of course give $D_Y^2 = -4$ on any surface.
This is therefore an appropriate choice for any surface with $h^{2,0} = 1$.  The
difference on $K3$ is that \emph{any} rational curve is a $(-2)$-curve.}  In
\sref{sec:toy_model} we will write down a toy model which implements the
setup we have described here.

\subsection{Flux and extra vector-like states}

In \sref{sec:unification}, we discussed an appealing scenario in which there
is an extra light vector-like pair of chiral multiplets in the
$(\rep{1},\rep{1},\pm 6)$ representation of $\GSM$, leading to F-theory
unification at $\sim 1.7\times 10^{18}\GeV$.  An important question to
answer is whether it is possible to get such a spectrum in these models.

Consider the simplest situation, where the extra states
arise on a particular component, $C$, of the $\rep{10}$ matter curve,
with the chiral families originating on another component (the calculation of
the spectrum does not always split up like this just because the matter
curve is reducible; see for example
\cite{Tatar:2009jk,Donagi:2011jy,Donagi:2011dv}).  Breaking down the
$\rep{10}{\oplus}\conjrep{10}$ under $\GSM$, the number of massless
fields in each representation is given by
\begin{equation*}
    \begin{array}{lcl}
    n_{(\conjrep{3},\rep{1},-4)} = h^0(C, \cL') &,&
         n_{(\rep{3},\rep{1},4)} = h^1(C, \cL')\\
     n_{(\rep{3},\rep{2},1)} = h^0(C, \cL'\otimes\cL_Y) &,&
         n_{(\conjrep{3},\rep{2},-1)} = h^1(C, \cL'\otimes\cL_Y) \\
     n_{(\rep{1},\rep{1},6)} = h^0(C, \cL'\otimes\cL_Y^2) &,&
         n_{(\rep{1},\rep{1},-6)} = h^1(C, \cL'\otimes\cL_Y^2) ~, \end{array}
\end{equation*}
for some common line bundle $\cL'$ \cite{Donagi:2008ca,Beasley:2008dc}.

For any line bundle $\cL$ on an algebraic curve $C$, we have the
Riemann-Roch formula:
\begin{equation*}
    h^0(C, \cL) - h^1(C, \cL) = \deg(\cL) + 1 - g_C ~.
\end{equation*}
We therefore see immediately that if we want extra $(\rep{1},\rep{1},\pm 6)$
states, but no others, then we must have $\deg(\cL_Y\rest{C}) = 0$, but
$\cL_Y\rest{C} \neq \cO_C$.  Obviously we then require $g_C > 0$, and
$\deg(\cL'\rest{C}) = g_C - 1$.  A simple way to achieve the desired outcome
is for $C$ to be an elliptic curve, and
$\cL'\rest{C} \cong \cL_Y\rest{C} \cong \cL$, where $\cL^{\otimes 3} = \cO_C$,
but $\cL \neq \cO_C$.  Here we outline one example of such a geometry,
without making any attempt to embed it in a consistent F-theory
compactification.

First, note that if $S \cong K3$ contains a non-singular elliptic curve $C$,
then it is in fact elliptically-fibred over $\IP^1$.\footnote{To prove this, start
with the short exact sequence
\begin{equation*}
    0 \longrightarrow \cO_S \longrightarrow \cO_S(C) \longrightarrow  \cO_S(C)\rest{C}
        \longrightarrow 0 ~,
\end{equation*}
and observe that the adjunction formula gives $\cO_S(C)\rest{C} \cong \cO_C$,
since both $K_C$ and $K_S$ are trivial.  Taking cohomology, the above
sequence then tells us that the linear system $|C|$ is one-dimensional.
It cannot have any base points, since $C\cdot C = 0$, so $S$ is
elliptically-fibred over $\IP^1$.}  Let us assume that it is a special elliptic
$K3$, which not only has a section, but has a non-trivial Mordell-Weil
group, with torsion subgroup $\IZ_3$ \cite{Shimada:2005ek}.  This means
that, as well as the zero section $\s$, there is another section $\s'$ such
that when restricted to a generic fibre, such as $C$, we have
$(\s' - \s)\rest{C} \nsim 0$, but $3(\s' - \s)\rest{C} \sim 0$.  Note that
$\s$ and $\s'$ are disjoint rational curves on $S$, so $\s' - \s$ can play
the role of $D_Y$, and the desired scenario arises if the $\rep{10}$ matter
curve contains a generic fibre $C$ as a component,
and $\cL'\rest{C} \cong \cL_Y\rest{C} \cong \cO_C(\s' - \s)$.

The  results above contradict \cite{Marsano:2009gv}, in which it was claimed
that any incomplete $SU(5)$ multiplets arising on curves threaded by
hypercharge flux satisfy $\d b_F = 0$ (where again
$\d b_F = 5\,\d b_1 - 3\,\d b_2 - 2\,\d b_3$).  In that work, expressions for
the $\d b_j$ are given in terms of the net chirality in each $\GSM$
representation; the implicit assumption is that the vector-like fields come in
complete $SU(5)$ multiplets, and so give no net contribution to $\d b_F$.
As has just been demonstrated, this is not generally true.

\newpage
\section{\texorpdfstring{$R$}{R}-symmetry} \label{sec:R_symmetry}

As mentioned earlier, perhaps the most obvious theoretical motivation
for Dirac gauginos is that they are necessary in a theory with an
unbroken (approximate, discrete) $R$-symmetry.  Conversely, Dirac
gauginos require some mechanism to suppress the usual Majorana
mass terms, and an unbroken $R$-symmetry is the most obvious way
to do this, at least in effective field theory.

In theories with extra dimensions, the unbroken supercharges come
from covariantly-constant spinors in the compact space, and
$R$-symmetries therefore arise from geometric symmetries which act
non-trivially on these spinors.  There is a slight subtlety here.  The
action of a geometric symmetry on tensorial quantities is always
well-defined, being given by the pushforward, whereas there is a sign
ambiguity in the action on spinorial quantities.  However, physics does
not care which sign we choose; we can see this from the fact that any
Lorentz-invariant Lagrangian must contain only terms with an even
number of spinors, so that an overall sign always cancels.  Alternatively,
note that a $2\pi$ rotation in the external dimensions is always a
symmetry, and this precisely changes the sign of all spinorial quantities.
These observations also show that $R$-parity should not really be
considered an $R$-symmetry \cite{Hall:1983id}: only symmetries of
order greater than two deserve this label.

To identify possible $R$-symmetries in the context of F-theory, first
assume that the Calabi--Yau fourfold $X$ is smooth.  Recall that if we
further reduce the theory to $(2+1)$ dimensions by compactifying one
space-like direction on a circle, the resulting theory is equivalent to
M-theory compactified on $X$.  The spin group in $(3+1)$ dimensions
is $SL(2,\IC)$, and $\cN=1$ supersymmetry is generated by a doublet
under this group, $Q_\a$.  Upon reduction to $(2+1)$ dimensions, the
spin group becomes $SL(2,\IR) \subset SL(2,\IC)$.  The doublet of
$SL(2,\IR)$ is real, so the real and imaginary parts of $Q_\a$ are now
independent, and we get $\cN=2$ SUSY in three dimensions.  From the
M-theory point of view, this comes about as follows.  A Calabi--Yau
fourfold has holonomy group $SU(4) \subset SO(8)$, and $SO(8)$ has
two eight-dimensional Majorana-Weyl spinor representations.  Under
$SU(4)$, one of these decomposes as
$\rep{8} = \rep{6}+\rep{1}+\rep{1}$ \cite{Slansky,Becker:1996gj}; the two
singlets represent the two real covariantly constant spinors, $\x_1$ and
$\x_2$, on $X$.  The complex spinor $\x = \x_1 + \ii\x_2$ then
corresponds to $Q_\a$.

In terms of $\x$, the holomorphic $(4,0)$ form on $X$ can be written as
${\O_X}_{ijkl} = \x^T\g_{ijkl}\x$, where $\g_{ijkl}$ is the anti-symmetric
product of the gamma matrices with holomorphic indices.  This makes it
easy to search for $R$-symmetries: we need automorphisms $g : X \to X$
such that $g_*\O_X \neq \O_X$.  For example, if $g^2 = \mathrm{id}_X$,
then we might have $g_*\O_X = -\O_X$.  We must therefore have
$g : \x \to \pm\ii\x$, where we are free to choose the sign, as discussed
above.  Such an automorphism of $X$ would thus correspond to a
$\IZ_4$ $R$-symmetry, but of a very restricted type: since
$g^2 = \mathrm{id}_X$, all superfields carry $R$-charge $0$ or $2$.  In
typical $R$-symmetric models, the quark and lepton superfields carry
$R$-charge $1$ (such that the fermionic components are neutral), so this
is not desirable.  Instead, we should consider an order-four symmetry $g$
which satisfies $g_*\O_X = -\O_X$.  This is again a $\IZ_4$ $R$-symmetry, but
now tensorial quantities can, in principle, carry any charge.  The
generalisation to other $\IZ_p$ is obvious.

In practice, of course, we are interested in singular fourfolds $X$, for which
the M-theory dual is defined on a crepant resolution $\widetilde X \to X$.
In general, there is no reason for $\widetilde X$ to share the symmetries of
$X$, but the F-theory limit is that in which all the resolution parameters
(which possibly break the symmetry) vanish.  So if $X$ is the limit of some
family of smooth fourfolds, all of which share a certain $R$-symmetry, then
by continuity we expect this $R$-symmetry to persist in the theory defined
on $X$, regardless of whether or not it admits a symmetric resolution.  We
will henceforth assume this to be true.

To detect $R$-symmetries we need an explicit representation of the
holomorphic $(4,0)$-form $\O_X$.  To get this in some generality, assume that
$X$ is given by a smooth Weierstrass model over $B$.  Let
$P = \IP\big(\cO_B{\oplus}K_B^{-2}{\oplus}K_B^{-3}\big)$, with
homogeneous coordinates $z, x, y$ on the fibres; then $X$ is given by the
vanishing of the Weierstrass polynomial $W = -y^2 z + x^3 + f x z^2 + g z^3$,
where $f$ and $g$ are sections of $K_B^{-4}$ and $K_B^{-6}$
respectively.  The adjunction formula for $X \subset P$ then leads to the
following short exact sequence \cite{GriffithsHarris}
\begin{equation} \label{eq:X_Poincare_res}
    0 \longrightarrow \O^5 P \longrightarrow \O^5 P(X) 
        \stackrel{\mathrm{P.R.}}{\longrightarrow} \O^4 X \longrightarrow 0 ~.
\end{equation}
The map labelled `P.R.' here is the Poincar\'e residue map, given by integrating
a $(5,0)$-form on $P$ over the boundary of an infinitesimal tubular
neighbourhood of $X$.  We now consider the long exact sequence in cohomology
following from the above.

The low-degree cohomology of the first term vanishes, which we can see as
follows: $B$ is the base of an elliptically-fibred Calabi--Yau, so $h^{p,0}(B) = 0$
for $p>0$.  Since $P$ is a $\IP^2$ bundle over $B$, and $h^{p,0}(\IP^2) = 0$ for
$p>0$, this implies by the Leray spectral sequence that $h^{p,0}(P) = 0$ for
$p>0$.  Since $P$ is K\"ahler, we have $h^{5,q}(P) = h^{5-q,0}(P)$, and hence
$H^0(P,\O^5 P) = H^1(P, \O^5 P) = 0$.

Putting the above results into the exact sequence in cohomology following from
\eqref{eq:X_Poincare_res}, we learn that the holomorphic $(4,0)$-form $\O_X$
is the Poincar\'e residue of the unique global section of $\O^5 P(X)$.  Explicitly, if
we let $\l$ be (the pullback to $P$ of) the unique holomorphic $(3,0)$-form on
$B$ with values in $K_B^{-1}$, then on the patch $y \neq 0$, we have
\begin{equation} \label{eq:OX_residue}
    \O_X = \oint_{W=0} \frac{y\, \l\wedge dx\wedge dz}{W} ~.
\end{equation}
The reader can check that the integrand is a well-defined meromorphic differential
form on $P$, and that it has no extra singularities at infinity in the
fibre (i.e.\! as $y \to 0$).

\subsection{\texorpdfstring{$R$}{R}-charge of the adjoint fields}\label{sec:adjoint_Rcharge}

$R$-symmetric Dirac masses require the adjoint chiral superfields to have
zero $R$-charge, which occurs if their wavefunctions on $S$ are invariant under
the geometric action of the $R$-symmetry.  This will typically need to be
checked on a case-by-case basis, but we will now see that some general
statements can be made when $h^{2,0}(S) = 1$, and explicit formulae are
available when $S \cong K3$.

When $h^{2,0}(S) = 1$, the unique holomorphic $(2,0)$-form $\O_S$ gives
rise to the massless adjoint fields.\footnote{For
$S \cong K3$, $\O_S$ will be everywhere non-zero; on other surfaces, it will
vanish along some curve.}  We have a short exact sequence corresponding to
$S \subset B$,
\begin{equation*}
    0 \longrightarrow \O^3 B \longrightarrow \O^3 B(S) 
        \stackrel{\mathrm{P.R.}}{\longrightarrow} \O^2 S \longrightarrow 0 ~.
\end{equation*}
Since $h^{3,0}(B) = h^{3,1}(B) = 0$ (again, because $B$ is the base of an elliptic
Calabi--Yau), and $h^{2,0}(S) = 1$ by assumption, we
learn that there exists a unique meromorphic $(3,0)$-form on $B$ with a pole
along $S$, and $\O_S$ is the Poincar\'e residue of this.

To go further, we need to specialise again, to the case where $S$ is a $K3$
surface, given by the vanishing of some section $s \in \G(B, K_B^{-1})$.
In this case, the global section of $\O^3 B(S)$ can be interpreted either as a
holomorphic $(3,0)$-form with values in $K_B^{-1}$, or a meromorphic
$(3,0)$-form with a pole along $S$; one is related to the other by dividing
by the section $s$.  This allows us to write an explicit formula for $\O_S$ in
terms of $\O_X$, as a double residue:
\begin{equation*}
    \O_S = \oint_{s=0} \frac{1}{s} \oint_{z=0} \frac{y \O_X}{x} ~.
\end{equation*}
This requires a little bit of explanation.  The section $B \subset X$ is given
globally by $z=0$, but $z$ actually has a third-order zero along $B$; locally,
it is $x$ which has a simple zero along $B$, giving the integrand here a
simple pole.  We see that after the first integral, we obtain a holomorphic
$(3,0)$-form on $B$ with values in $K_B^{-1}$, since $\O_X$ is a section of
the trivial line bundle $K_X$, and $y/x \sim K_B^{-1}$.

We now see that in order to obtain adjoint fields with zero $R$-charge, we need to
choose a section $s \in \G(B, K_B^{-1})$ which transforms with the same
charge as $y\O_X/x$ under the $R$-symmetry, so that $\O_S$ is invariant.

When $S$ is not a $K3$ surface, we cannot write down a general relationship
between $\O_S$ and $\O_X$, because the global section of $\O^3B(S)$ can
no longer be written down in terms of $\O_X$.  It should still be possible to find
explicit expressions in most cases, and therefore find the $R$-charge of the
adjoint fields, but we cannot find a \emph{general} formula.

\section{A toy example}\label{sec:toy_model}

It is relatively easy to write down a geometry which realises many of the
features discussed in this paper (I make no attempt at engineering a
realistic matter sector).

Let $B = \IP^3$, which is arguably the simplest threefold base we could
use, with homogeneous coordinates $(u_0, u_1, u_2, u_3)$.  Then the
ambient fivefold is the projective bundle
$\IP\big(\cO_B{\oplus}\cO_B(8){\oplus}\cO_B(12)\big)$, with
homogeneous coordinates $(z, x, y)$ on the fibres, in which our Calabi--Yau
fourfold $X$ is given by the vanishing of the generalised Weierstrass
polynomial
\begin{equation}\label{eq:generalised_Weierstrass}
    W = -y^2 z -a_1 xyz - a_3 yz^2 + x^3 + a_2 x^2 z + a_4 xz^2 + a_6z^3 ~,
\end{equation}
where $a_k$ is a homogeneous polynomial of degree $4k$ in the $u_m$.

When $X$ is smooth, we can specialise \eqref{eq:OX_residue} to the case at
hand to get a residue formula for the holomorphic $(4,0)$ form:
\begin{equation*}
    \O_X = \oint_{W=0} \frac{y\, u_0\, du_1\wedge du_2 \wedge du_3 \wedge dx \wedge dz}{W} ~.
\end{equation*}
As explained in \sref{sec:R_symmetry}, a potentially realistic $R$-symmetry
can be obtained from an order-four automorphism $g : X \to X$ under which
$\O_X \to -\O_X$.  There is a simple choice here which achieves this:
\begin{equation}\label{eq:Z4_action}
    g_4 : u_m \to \ii^m u_m ~,
\end{equation}
which extends to an order-four symmetry of the fourfold $X$ if we choose the
$a_k$ to be invariant.  It is easy to check that even with this restriction, $X$ is
generically smooth.

The next step is to specify a $K3$ surface $S$ inside $B \cong \IP^3$, to play
the role of the GUT brane.  Any quartic polynomial $s$ in $\IP^3$ defines a
$K3$, but a generic quartic hypersurface will have Picard number equal to
one, corresponding to the hyperplane class inherited from $B$, and therefore
we will be unable to turn on hypercharge flux.  So we must choose a special
family of quartics.  Note that any smooth hyperplane section, $C$, will
have self-intersection $4$ in $S$.  By the formula \eqref{eq:genus_formula},
$C$ will be a curve of genus 3.  One way that $S$ can have extra divisor
classes is if some of these hyperplane sections become reducible, i.e., split
into a union of lower-genus curves.

In fact, it is convenient to take a slightly different point of view.  In
\sref{sec:flux}, we showed that if $C_1$ and $C_2$ are disjoint rational
curves in $S$, which are homologous in $B$, then turning on hypercharge
flux along $D_Y = C_1 - C_2$ satisfies the criteria for a massless
hypercharge gauge boson and no exotic charged states coming from the
adjoint of $SU(5)$.  So define two disjoint, homologous, rational curves in
$B$:
\begin{equation*}
    C_1 = \{u_0 = u_1 = 0\} ~,~ C_2 = \{u_2 = u_3 = 0\} ~,
\end{equation*}
and now consider only those quartic hypersurfaces which contain both
$C_1$ and $C_2$.\footnote{To make contact between the two approaches,
note that $S$ containing, say, $C_1$, is equivalent to the hyperplane
section $u_0 = 0$ splitting into the union of $C_1$ and some degree-three
curve.}  As explained in \sref{sec:adjoint_Rcharge}, we must also choose
our quartic polynomial $s$ such that $s \to -s$ under the action of $g_4$, so
that the adjoint chiral superfields will have $R$-charge 0.
Note also that the divisor class $C_1 - C_2$ is invariant under $g_4$, so
the hypercharge flux preserves the $R$-symmetry.

An explicit example, which is readily checked to be smooth, is given by
\begin{equation*}
    s = u_0 u_2^3 + u_0^2 u_3^2 + u_0^3 u_2 + u_1 u_3^3 + u_1^2 u_2^2 + u_1^3 u_3 ~.
\end{equation*}

Finally, to specialise to those $X$ which have an $SU(5)$ singularity along
$S$, we must take the coefficients in \eqref{eq:generalised_Weierstrass} to be
$a_k = s^{k-1} q_k$, where each $q_k$ is a quartic polynomial, transforming
as $q_k \to (-1)^{k+1}q_k$ under the action of $g_4$, and none of them are
equal to $s$.  It is easy to check that there is enough freedom
that no extra singularities necessarily occur.

The family of Calabi--Yau fourfolds constructed in this section can be used
as the basis for a family of supersymmetric F-theory models with an unbroken
$\IZ_4$ $R$-symmetry, in which $SU(5)$ is broken by hypercharge flux in such
a way that the $U(1)_Y$ gauge boson remains massless, and the only
massless chiral fields descending from the adjoint of $SU(5)$ are those which
fill out the adjoint of the standard model gauge group, which moreover have
$R$-charge $0$.  This is a promising
starting point for a viable SUSY model with Dirac gauginos, but there is
obviously a lot more work to do to write down a complete, consistent model,
and this will be deferred to future work.  It may be that the particular fourfolds
here are too simple for realistic model-building, but the construction illustrates
the general ideas in a clear way.

\vspace{-2ex}
\section{Conclusions} \label{sec:conclusion}

In this paper I have begun the study of F-theory GUT models with Dirac
gauginos.  In particular, the conditions under which the requisite massless
chiral adjoints arise, but `off-diagonal' components of the $SU(5)$ adjoint
are absent, were shown to be easily satisfied.  I also showed explicitly, in
the case where the visible sector resides on a $K3$ surface, how to
engineer an $R$-symmetry under which these have the correct charge;
this is potentially an important ingredient in these models, as it suppresses
Majorana gaugino masses.

I have said very little about the matter sector, except to indicate how one
might arrange for the presence of the light vector-like pairs required to
obtain a realistic GUT scale, and nothing at all about how to break
supersymmetry in a realistic way (although see \aref{app:Dirac_masses}
for a telegraphic account of the field-theoretic considerations).  These are
obviously the most important next steps in developing quasi-realistic
F-theory models with Dirac gauginos.

My over-arching point is the following.  Given the strong bounds which the
LHC has already set on MSSM-like theories, it is important to consider
alternative scenarios if we wish to retain supersymmetry as a solution to the
hierarchy problem.  Models with Dirac gauginos are one compelling option,
and it is surprising that they have basically not yet been considered by the
string phenomenology community.  While this work has taken only rudimentary
steps, I hope that it will spark some interest in the subject.

\subsection*{Acknowledgements}

I would like to thank Ron Donagi for helpful conversations, and Matthew
McCullough for comments on a draft version of the paper.  This work was
supported by the Engineering and Physical Sciences Research Council
[grant number EP/H02672X/1].

\appendix

\section{Generating Dirac gaugino masses} \label{app:Dirac_masses}

Arranging for the low-energy theory to contain adjoint chiral multiplets is a
necessary condition to have Dirac gauginos, but we must also ensure that a
large Dirac mass term is generated after SUSY breaking.  This appendix
contains no original work, but is included to make the paper more
self-contained, and to point out some of the difficulties which will need to be
overcome to build realistic F-theory models with Dirac gauginos.

Denote an $SU(n)$ adjoint chiral multiplet by
$\mb{\Phi} = \phi + \sqrt{2}\,\th\psi + \ldots$, and the field strength superfield
by $\mb{W_\a} = -\ii\l_\a + \th_\a D + \ldots$, with gauge indices suppressed.
Dirac gaugino masses arise most simply via an interaction with a hidden
sector $U(1)$ gauge field which obtains a $D$-term VEV.  In terms of the
field strength of this hidden $U(1)$, $\mb{W}'_\a = -\ii\l'_\a + \th_\a D' + \ldots$,
the term we need is\footnote{Dirac masses can also be obtained from the
$F$-term VEV of a chiral field $X$, via operators like
$\frac{1}{\widetilde M^3}\int d^4\th\, X^\dagger X \Tr(W^\a \cD_\a\Phi)$, but
these are typically suppressed by an extra factor of $\frac{F}{M^2}$ relative
to other soft masses (although see \cite{Abel:2011dc} for an example where
this is not the case).}
\begin{equation}\label{eq:Dirac_mass}
    \cL_D = \frac{\sqrt{2}\tilde y}{M} \int d^2\th\, \mb{W'^\a \Tr (W_\a \Phi)} + \mathrm{H.c.} = 
        \frac{\ii \tilde y}{M} D'\,\Tr(\l \psi) + \ldots ~,
\end{equation}
where $M$ is some mass scale, and $\tilde y$ some dimensionless
constant.

\begin{figure}[ht]
\begin{center}
    \includegraphics[width=.6\textwidth]{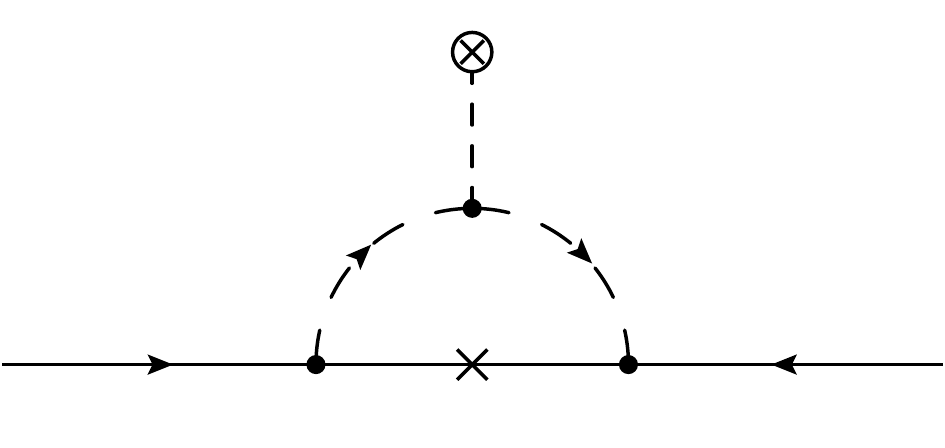}\\
    \place{1.9}{0.4}{$\l$}
    \place{4.5}{0.4}{$\psi$}
    \place{3.4}{1.8}{$D'$}
    \place{2.3}{1.1}{$C/C'$}
    \place{3.8}{1.1}{$C/C'$}
    \vspace{-4ex}
        \parbox{\textwidth}{
        \caption{\label{fig:Dirac_loop}
        \small
        The one-loop diagram which generates \eqref{eq:Dirac_mass}.  The
        crossed line represents the propagator for the Dirac fermion consisting
        of the superpartners of $C$ and $C'$.  The diagrams in which the scalar
        is respectively $C$ and $C'$ add constructively.
        }}
\end{center}
\end{figure}

To generate \eqref{eq:Dirac_mass}, we can introduce a vector-like pair of
chiral fields $\mb{C, C'}$, in the bi-fundamental representation $(\rep{n}, 1)$
and its conjugate, respectively, and take their superpotential couplings to
be
\begin{equation*}
        W_C = M \mb{C C'} + y\, \mb{C' \Phi C} ~.
\end{equation*}
This leads to the generation of \eqref{eq:Dirac_mass} via the one-loop diagram
shown in \fref{fig:Dirac_loop}.  Note that this simple model is $R$-symmetric if
we assign an $R$-charge of $1$ to the fields $\mb{C, C'}$, and $0$ to the
chiral adjoint superfield $\mb{\Phi}$ (the latter is required for the Dirac gaugino
masses to be $R$-symmetric).

Unfortunately, the simple model presented here also generates a holomorphic
mass term for the adjoint scalars, leading to a tachyonic mass for one
component, and thus the breaking of colour $SU(3)$, for example.  This
problem can be solved by taking more than one pair of messengers, and
imposing certain conditions on their couplings to the adjoint fields
\cite{Benakli:2008pg}, but this does pose an extra model-building challenge.

Pure $D$-term breaking of SUSY gives problematically-light sleptons,
which led the authors of \cite{Benakli:2010gi} to consider combined $F$- and
$D$-term SUSY breaking, of the same order.  This can easily be achieved in
explicit models \cite{Dumitrescu:2010ca}.

Finally, we note that the gauginos will obtain unavoidable Majorana
masses from anomaly-mediation, of order
$\frac{g^2}{16\pi^2}m_{3/2}$ (where $m_{3/2}$ is the gravitino mass), which
we should ensure are at least an order of magnitude or two smaller than the
Dirac masses, lest the benefits conferred by Dirac gauginos be lost.

\newpage

\bibliographystyle{utphys}
\bibliography{references}

\end{document}

%% file: macros.tex
\usepackage{amssymb,amsmath}

\renewcommand{\a}{\alpha}

\newcommand{\g}{\gamma}\newcommand{\G}{\Gamma}
\renewcommand{\d}{\delta}

\renewcommand{\th}{\theta}

\renewcommand{\l}{\lambda}\renewcommand{\L}{\Lambda}
\newcommand{\m}{\mu}

\newcommand{\x}{\xi}

\newcommand{\p}{\pi}

\newcommand{\s}{\sigma}

\newcommand{\vph}{\varphi}

\renewcommand{\o}{\omega}\renewcommand{\O}{\Omega}

\newcommand{\cD}{\mathcal{D}}

\newcommand{\cL}{\mathcal{L}}

\newcommand{\cN}{\mathcal{N}}
\newcommand{\cO}{\mathcal{O}}

\newcommand{\IC}{\mathbb{C}}

\newcommand{\IP}{\mathbb{P}}

\newcommand{\IR}{\mathbb{R}}

\newcommand{\IZ}{\mathbb{Z}}

\newcommand{\rep}[1]{\mathbf{#1}}
\newcommand{\conjrep}[1]{\overline{\mathbf{#1}}}

\newcommand{\quotient}[1]{_{\hskip-2pt\lower1pt\hbox{$/$}\lower2pt\hbox{\hskip-1pt$#1$}}}
\newcommand{\ii}{\text{i}}

\def\place#1#2#3{\vbox to0pt{\kern-\parskip\kern-7pt
                             \kern-#2truein\hbox{\kern#1truein #3}
                             \vss}\nointerlineskip}

\newcommand{\mb}[1]{\mathbf{#1}}
\newcommand{\rest}[1]{\big\vert_{#1}}

\newcommand{\sref}[1]{section~\ref{#1}}
\newcommand{\aref}[1]{appendix~\ref{#1}}
\newcommand{\fref}[1]{Figure~\ref{#1}}
\newcommand{\tref}[1]{Table~\ref{#1}}

\newcommand{\diag}{\mathrm{diag}}

\newcommand{\GeV}{~\mathrm{GeV}}
\newcommand{\GSM}{G_{\mathrm{SM}}}
\newcommand{\GUT}{\mathrm{GUT}}

\newcommand{\TeV}{~\mathrm{TeV}}
\newcommand{\Tr}{\mathrm{Tr}}
\newcommand{\Vol}{\mathrm{Vol}}

\newcommand{\symm}[1]{_{\hskip-3pt\lower3pt\hbox{$\left\{#1\right\}$}}}

\newcommand{\cicystop}{~\lower8pt\hbox{.}}

\def\place#1#2#3{\vbox to0pt{\kern-\parskip\kern-7pt
                             \kern-#2truein\hbox{\kern#1truein #3}
                             \vss}\nointerlineskip}

\newcommand{\beq}{\begin{equation}}
\newcommand{\eeq}{\end{equation}}
\newcommand{\bea}{\begin{eqnarray}}
\newcommand{\eea}{\end{eqnarray}}
\newcommand{\bean}{\begin{eqnarray*}}
\newcommand{\eean}{\end{eqnarray*}}

\newcommand{\comment}[1]{}

\newcommand{\+}{\phantom{-}}
\newcommand{\id}{{\bf 1}}

